\documentclass[twocolumn,amssymb,fleqn,showpacs]{revtex4} 
\usepackage{epsfig,amssymb,amsmath,graphicx,subfigure} 

\newcommand{\be}{\begin{eqnarray}}
\newcommand{\ee}{\end{eqnarray}}

\begin{document}

\title{Self-Propulsion of Droplets by Spatially-Varying Roughness}
\author{Zhenwei Yao and Mark J. Bowick}
\affiliation{Physics Department, Syracuse University, Syracuse,
New York 13244-1130, USA}
\begin{abstract}

Under partial wetting  conditions, making a substrate uniformly
rougher enhances the wetting characteristics of the corresponding
smooth substrate {--} hydrophilic systems become even more
hydrophilic and hydrophobic systems even more hydrophobic. Here we
show that spatial texturing of the roughness may lead to
spontaneous propulsion of droplets. Individual droplets are driven
toward regions of maximal roughness for intrinsically hydrophilic
systems and toward regions of minimal roughness for intrinsically
hydrophobic systems. Spatial texturing can be achieved by
wrinkling the substrate with sinusoidal grooves whose wavelength
varies in one direction (inhomogeneous wrinkling) or
lithographically etching a radial pattern of fractal (Koch curve)
grooves on the substrate.  Richer energy landscapes for droplet
trajectories can be designed by combining roughness texturing with
chemical or material patterning of the substrate.

\end{abstract}
 \maketitle

Consider a liquid droplet partially wetting a solid substrate such
as glass in contact with a gas such as air. Broadly speaking a
substrate may wet easily (hydrophilic) or poorly (hydrophobic)
depending on the nature of the substrate, the liquid and the gas.
More specifically the three relevant interfacial surface tensions
determine the contact angle made by the liquid-gas contact line
meeting the plane of the substrate.  The contact angle is less
than $90^{\circ}$ for hydrophilic systems and greater than
$90^{\circ}$ for hydrophobic systems.  A totally wetting thin film
corresponds to vanishing contact angle and a complete spherical
drop balanced at a point on the substrate corresponds to the
superhydrophobic limit with a $180^{\circ}$ contact angle.

Uniform surface roughness amplifies the basic wetting
characteristics of the corresponding planar system. For
hydrophilic/hydrophobic systems the greater substrate area, for a
given planar projection, available on the rough substrate makes
the wetting more/less favorable and lowers/raises the contact
angle. What about surfaces with spatially inhomogeneous
properties? Although variable chemical patterning \cite{chem} and
Leidenfrost droplets contacting hot surfaces with asymmetric
sawtooth patterns \cite{Q_hot, L_hot} have been thoroughly
explored, this letter addresses the energetic driving forces
acting on droplets on a substrate with pure {\em inhomogeneous}
roughness and no other variability.  Droplets will spontaneously
move around in the landscape of the surface topography maximizing
or minimizing the roughness for naturally hydrophilic/hydrophobic
systems respectively.  Thus self-propelled droplets can be
engineered to follow prescribed paths without external drive by
appropriately designing the surface topography.

\begin{figure}[h]  
\subfigure[]{
\includegraphics[width=1.5in]{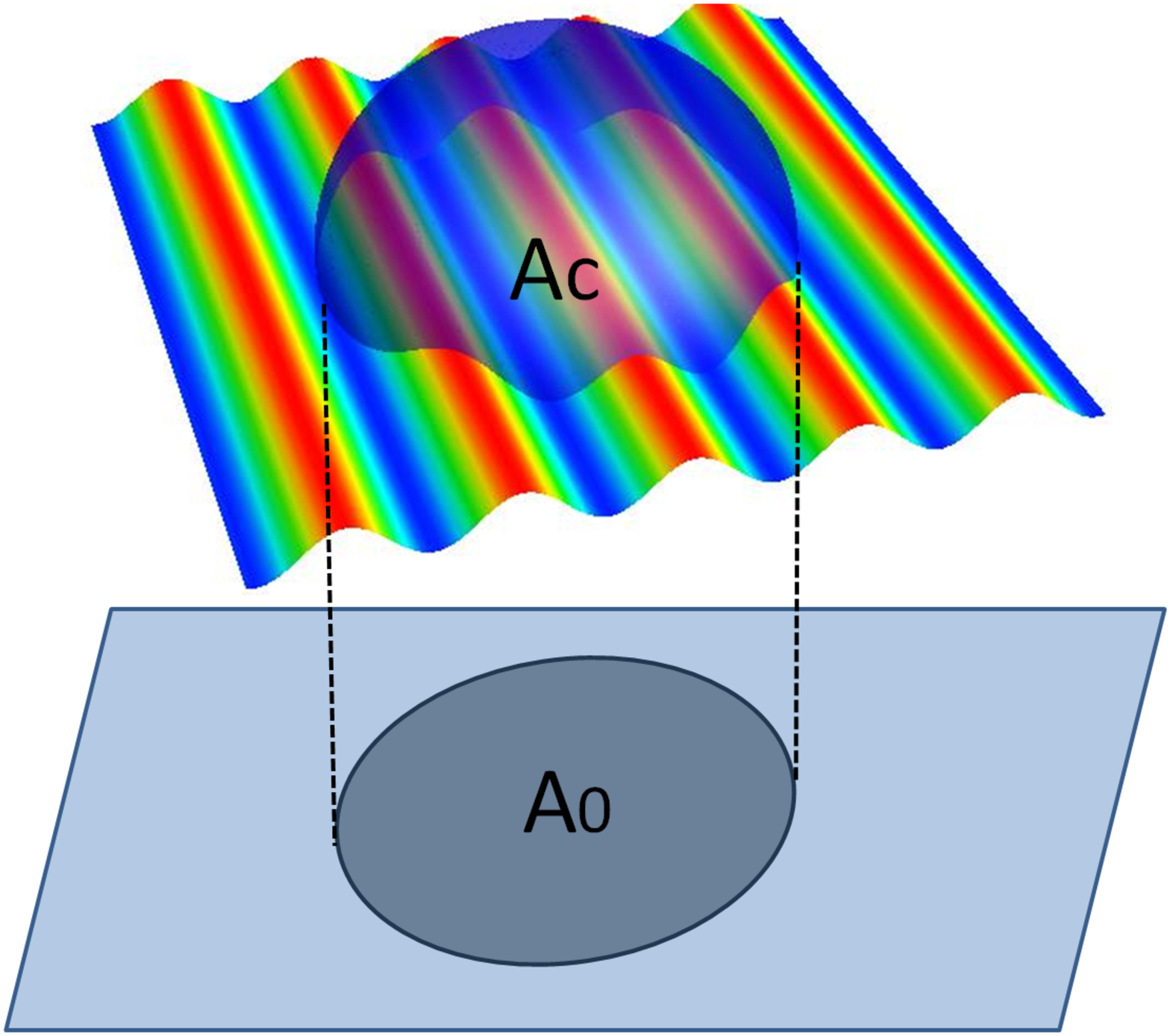}}
\hspace{-0.1in} \subfigure[]{
\includegraphics[width=1.6in,bb=0 100 560 344]{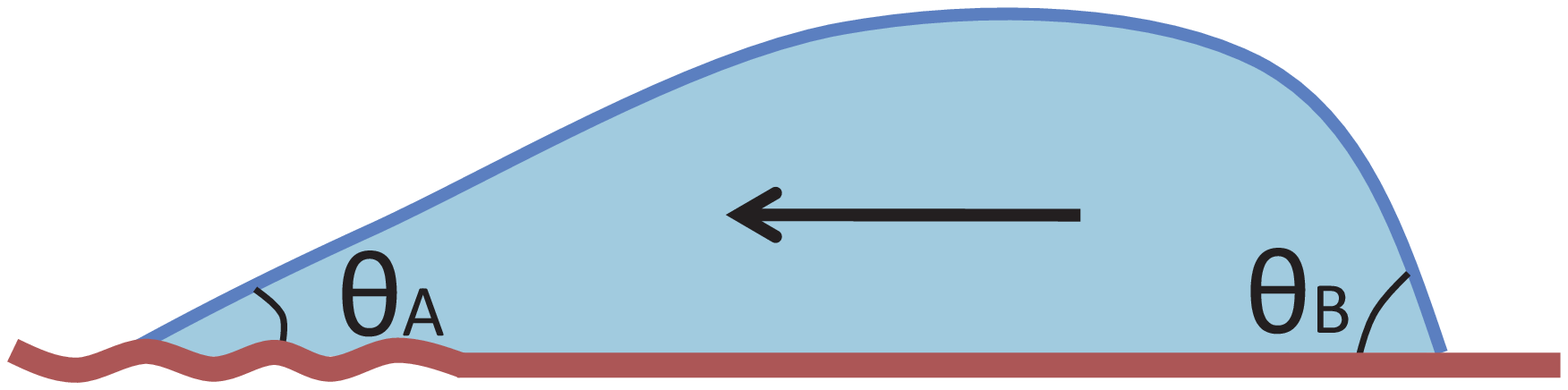}}  
\caption{(a) A liquid droplet sitting  on a rough substrate has
more contact area  with the substrate ($A_c>A_0$) than the same
droplet on an otherwise identical flat substrate. (b) A liquid
droplet partially wetting a hydrophilic substrate with
inhomogeneous roughness has a smaller contact angle at the rougher
end ($\theta_A$) than at the smoother end ($\theta_B$). }
\label{exp1}
\end{figure}

The principle of driving liquid droplets via roughness gradients
is simple. The free energy of a liquid droplet on a substrate is
$F=-I A_c + \sigma_{SV}A_t+\sigma_{LV}A_{LV}$, where
$I=\sigma_{SV}-\sigma_{SL}$ is the imbibition parameter
\cite{deGennes}, $\sigma_{SL}$, $\sigma_{SV}$ and $\sigma_{LV}$
are the respective surface tensions between the three phases
(Solid/Liquid/Vapor), $A_c$ is the contact area between the
droplet and substrate, $A_t$ is the (constant) total area of a
substrate and $A_{LV}$ is the area of the liquid-vapor interface,
which is taken to be constant even when a droplet moves. The
system of a liquid droplet on a rough substrate may also be viewed
as a droplet on a flat substrate with an effective imbibition
parameter $I_{eff}$ resulting from the roughness. $I_{eff}$ is
defined by $I A_c \equiv I_{eff} A_0$, where $A_0$ is the planar
projection of the actual contact area. For rough surfaces,
$A_c>A_0$, as shown in Fig.\ref{exp1}(a), and therefore
$I_{eff}/I>1$. Up to irrelevant constants, the free energy of a
droplet on a rough substrate is \be F=-I_{eff}(\vec{x}) A_0
,\label{free_energy}\ee where $I_{eff}$ varies from place to place
when the spatial roughness is inhomogeneous. The wetting
characteristics of a substrate/liquid composite system determines
the sign of $I$ and therefore $I_{eff}$. A hydrophilic system is
characterized by $I >0$ and an acute contact angle $\theta=
\arccos(I/\sigma_{LV})$. A hydrophobic system is characterized by
$I<0$ and an obtuse contact angle \cite{deGennes}. When $I$, and
so $I_{eff}$, is positive (negative) a substrate lowers
(increases) its free energy when covered by a liquid. This
spontaneously drives droplets on hydrophilic (hydrophobic)
substrates towards rougher (smoother) regions respectively.
Eq.(\ref{free_energy}) can also be used to understand the movement
of droplets on a chemically heterogeneous substrates where
$I_{eff}(\vec{x})$ depends on the wetting characteristics of the
chemical composition at the corresponding position on the
substrate.

The self-propulsion of liquid  droplets on substrates with
inhomogeneous roughness can also be understood in terms of the
uneven distribution of the Laplace pressure across the droplet.
The contact angle for a rough substrate ($\theta_r$) is given by
$\cos\theta_r=r \cos\theta$, with $r=A_c/A_0$ \cite{Wenzel}. Thus
surface roughness amplifies the intrinsic wetting properties of
the corresponding planar substrate. Take a droplet spanning a
hydrophilic surface that is rougher on the left than on the right,
as illustrated in Fig.\ref{exp1} (b). The contact angle is then
smaller on the left than on the right: $\theta_A < \theta_B$. The
mean curvature $H$ at the B end thus exceeds that at the A end,
leading to a Laplace over-pressure ($P=2 \sigma_{LV} H$) gradient
from right to left driving the droplet to the rougher part of the
surface. The reverse argument applies to a hydrophobic substrate,
leading to motion towards the smoother part of the substrate. The
self-propulsion of a droplet on a substrate with spatially varying
roughness clearly requires the size of the contact disk between
the droplet and the substrate to be larger than the typical size
over which the roughness varies significantly.

\begin{figure}
\centering \subfigure[]{
\includegraphics[width=1.32in, bb=157 21 522 473]{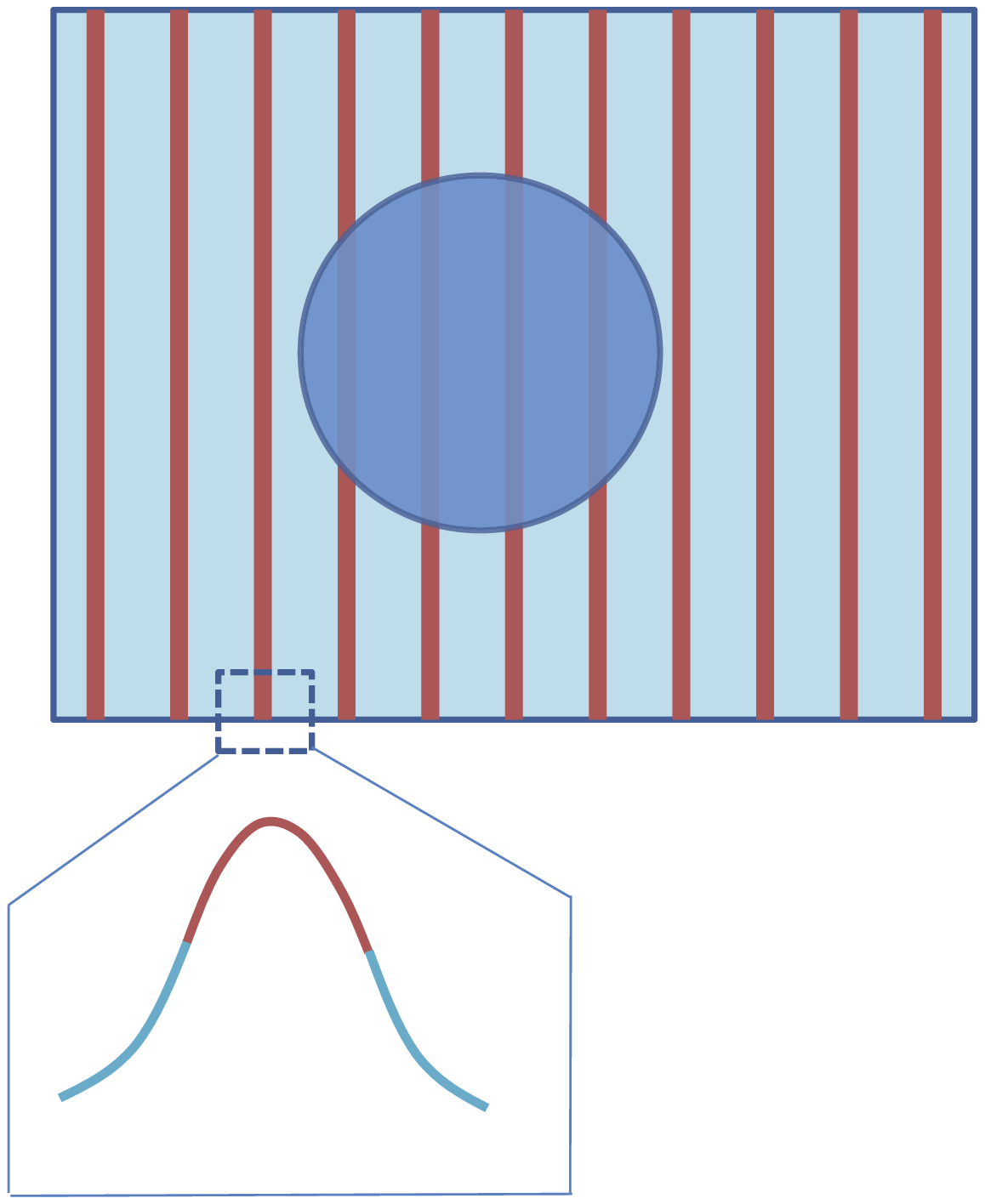}}
\hspace{-0.01in} \subfigure[]{
\includegraphics[width=1.92in]{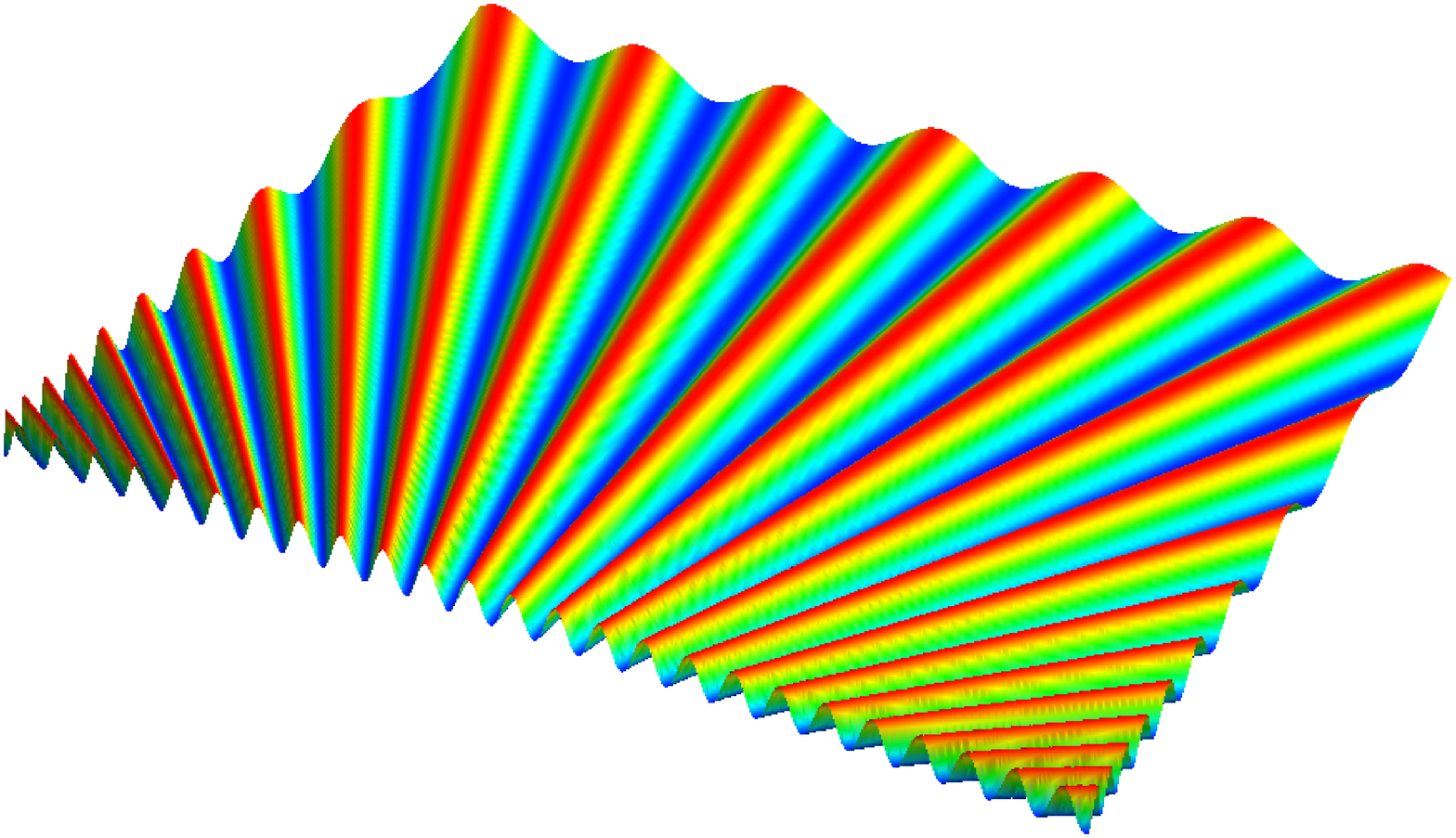}} 
\caption{(a) A liquid droplet partially wetting a substrate with a
uniaxial sinusoidally modulated roughness. (b) Schematic plot of
sinusoidal grooves with wavenumber monotonically increasing in the
direction orthogonal to the sinusoidal height profile. }
\label{sinusoidal2}
\end{figure}

\begin{figure}[h]
\centering
\includegraphics[width=2.2in, bb=0 0 309 216]{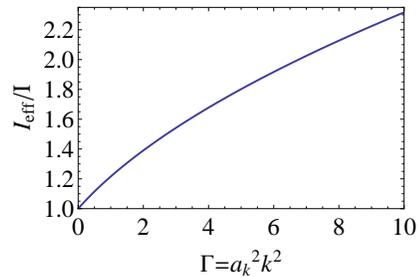}
\caption{The ratio of the effective imbibition parameter to the
physical imbibition parameter versus $\Gamma=a_k^2 k^2$  for a
liquid droplet partially wetting a sinusoidally modulated substrate. } \label{eff_sigma}
\end{figure}

To be specific, consider a droplet  on a uniaxial sinusoidal
substrate, as shown in Fig.\ref{sinusoidal2}(a), realizable via
wrinkled membranes \cite{wrinkles}. The height of the substrate is
represented by $z(x)=a_k \cos(k x)$, with translational invariance
along the y-direction. The roughness amplitude $a_k$ will be taken
much smaller than the maximum height of the droplet so that the
shape of the liquid-air interface is unaffected by the shape of
the substrate. For small amplitude roughness liquid droplets can
be in complete contact with the substrate, since air pockets do
not form underneath the liquid  \cite{Wenzel, deGennes}. The
effective imbibition parameter is given by \be
\frac{I_{eff}}{I}=\frac{4}{\pi} \int_0^1 dy \sqrt{1-y^2}
\sqrt{1+\Gamma \sin^2(\tilde{k} y)} ,\label{sigmaeff_sin1}\ee
where $\tilde{k}=kR$ is the dimensionless wavenumber and $\Gamma =
a_k^2 k^2$. Clearly $\Gamma$, arising from the gradient of the
substrate height, is the parameter controlling the effective
imbibition. For $R>>\frac{2\pi}{k}$, Eq.(2) simplifies to
\cite{cal_1} \be \frac{I_{eff}}{I}=\frac{2}{\pi} R(\Gamma)
,\label{sigmaeff_sin2}\ee where $R(x)=\int_0^{\pi/2}d\theta
\sqrt{1+x\sin^2\theta} $. In this limit the effective imbibition
parameter is dependent only on the product of the amplitude and
the wavenumber of the sinusoidal substrate. Fig.\ref{eff_sigma} is
a numerical plot of the monotonic growth of $I_{eff}/I$ vs.
$\Gamma$. $I_{eff}/I$ is doubled for $\Gamma\approx 8$. A simple
gradient of the effective imbibition parameter over the substrate
can be achieved by varying the wavenumber $k$ along the groove
axis (y), as shown in the schematic Fig.\ref{sinusoidal2}(b).
Droplets will migrate to maximize/minimize the contact area for
intrinsically hydrophilic/hydrophobic substrates. The magnitude of
the driving force along the groove axis is proportional to the
gradient of the effective imbibition parameter: \be \nabla_y
I_{eff} =\frac{2 I }{\pi} [R(\Gamma)-S(\Gamma)] \frac{d\ln
\tilde{k}(y)}{dy} ,\ee where $S(x)=\int_0^{\pi/2}\frac{d\theta
}{\sqrt{1+x \sin^2\theta}}$, and $R(x)-S(x)= \frac{\pi
x}{4}-\frac{3 \pi
   x^2}{32}+{\cal O} \left(x^3\right)$. The driving force thus
depends rather weakly on $\tilde{k}(y)$ and vanishes as $\Gamma$
approaches zero.

\begin{figure}[h]
\centering
\includegraphics[width=2.4in, bb=14 79 524 248]{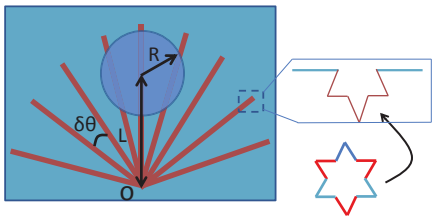}  
\caption{A substrate etched by fractal grooves. The
cross-sectional shape of the grooves is the lower half of the Koch
curve, as shown in the inset. } \label{exp2}
\end{figure}

Rough substrates may also be designed based on fractals
\cite{Kao}. Consider a substrate with etched grooves whose cross
section is the lower half of the Koch curve, as illustrated in
Fig.\ref{exp2} \cite{Koch}. The Koch curve may be constructed by
starting with an equilateral triangle of side length $a_0$ and
perimeter $L_0=3 a_0$, then recursively adding equilateral
triangles symmetrically on each line segment. After $n$ steps, the
perimeter of the new graph becomes $L_n= L_0 (4/3)^n $ with the
length of each elementary line segment being $a_n=a_0 (1/3)^n$.
The Hausdorff, or self-similarity, dimension of the Koch curve is
$d_H=\ln 4 / \ln 3 \approx 1.26$ \cite{fractal_book}. Now consider
a set of close-packed evenly aligned straight grooves constructed
from an $n$-th order Koch curve. The distance between two
neighboring grooves is twice the breadth of a groove. The contact
area between a droplet of radius $R$ and the fractal substrate, in
the limit $R$ much bigger than the breadth of a groove, is
$A_c=\pi R^2 [1+ q(n) ] $, where $q(n=0)=3/4$ and $q(n>0)=
2(4/3)^{n-2}-1/2$. Since $q(n)$ is always positive, $A_c/A_0>1$.
The effective imbibition parameter is \be \frac{I_{eff}}{I}=1+q(n)
.\label{eff_frac} \ee Thus the effective imbibition parameter
depends only on the order of the Koch curve $n$ and is independent
of the seed side length $a_0$. $I_{eff}/I$ is two for $n=1$ and
exceeds $10$ for $n=8$. Note that the extra volume of liquid
inside the finer structure of the fractal grooves is negligible in
the limit of large $n$ as the area $A_n$ of the $n$-th order Koch
curve converges: $\textrm{lim}_{n\rightarrow \infty}
A_n=2\sqrt{3}a_0^2/5 \sim a_0^2 << a_0 h$, where $h$ is the
droplet thickness.

An effective roughness gradient  can be made by etching a radial
array of grooves with Koch cross-section, as sketched in
Fig.\ref{exp2}. We take an intrinsically hydrophobic surface
inclined at an angle $\alpha$ to the horizontal. A droplet sitting
near the origin (O) of the radial array will move upward to reduce
the contact energy provided the gradient of the effective
imbibition parameter is sufficient to overcome gravity. The number
of grooves covered by a droplet of size $R$ is
$N=2R/(L\delta\theta)$, where $\delta \theta$ is the angular
distance between neighboring grooves and $L$ is the distance of
the droplet from the origin. For simplicity, consider the
$\lq\lq$far-field" limit in which the droplet is sufficiently far
from the origin that all the grooves under the droplet are
effectively parallel. The change in contact area due to the
roughness is $ \Delta A=A_c-A_0= \frac{1}{2}N \pi (b_n-l_0)R $,
where $b_n$ is the area of a groove of unit length with
$b_n(n>0)=\frac{3a_0}{2}(\frac{4}{3})^n$ and $b_n(n=0)=5a_0/3$,
and $l_0=2a_0/3$ is the groove breadth. This results in an
effective imbibition parameter \be
\frac{I_{eff}}{I}=1+\frac{b_n-l_0}{L
\delta\theta}.\label{sigma_Koch_uphill} \ee

A droplet rising up a distance $\delta L$ increases its
gravitational potential energy by $\delta W=mg \delta L
\sin\alpha$, where $m \sim \rho R^2 h$ is the mass of the droplet.
Meanwhile the surface energy decreases by $\delta F= A_0 \delta
I_{eff}$, where $\delta I_{eff}=I_{eff}(L+\delta L)-I_{eff}(L)$.
Spontaneous climb therefore requires $\delta I_{eff}/\delta L >
mg\sin\alpha/A_0$. Inserting Eq.(\ref{sigma_Koch_uphill}) for the
effective imbibition parameter yields \be \frac{b_n-l_0}{l_0}
> \frac{L^2 \delta \theta\ mg\sin\alpha}{A_0I l_0}
.\label{uphill_cond} \ee This condition can be satisfied for large
droplets near the origin on substrates with dense grooves. The
right hand side of Eq.(\ref{uphill_cond}) is of order $10
\sin\alpha$ for $L\sim \textrm{cm}, R\sim \textrm{mm}, I \sim
\textrm{mN/m}, \delta\theta \sim a_0/L$ and $h \sim 0.1 R$. Since
$(b_n-l_0)/l_0\sim 10$ for $n=5$, radially carved grooves made of
the $5$-th order Koch curves would generate sufficient roughness
gradient to drive droplets uphill.

There are several points in our analysis  that may ultimately call
for a more thorough treatment. Sharp substrate edges impede the
motion of droplets via pinning of the triple line \cite{NB}.
Adhesion hysteresis may also arise from the microscopic
interactions between a droplet and the substrate \cite{NN}. These
two effects are the main source of frictional energy dissipation
\cite{Nosonovsky}. We have neglected entirely viscous dissipation
due to internal fluid flow within moving droplets
\cite{deGennes,Landau}.

Droplet flow driven by  inhomogeneous surface roughness is
strictly downhill according to the gradient of the height profile
of the surface but one may also vary the chemical composition of
the surface so that the intrinsic surface tensions are spatially
dependent. The combination of chemical and roughness patterning
offers a rich variety of potential structures to obtain desired
flow patterns. The African beetle \textit{Stenocara} fog-basks by
tilting forward into the early morning fog-laden wind of the Namib
desert and collecting micron-sized water droplets on the smooth
hydrophilic peaks of its fused overwings (elytra) \cite{PL:2001}.
Once a sufficiently massive droplet is formed it rolls downhill
against the wind to pool in textured hydrophobic waxy troughs and
from there to the beetle's mouth. Surface structures modeled on the Stenocara wings
have been synthesized by creating hydrophilic patterns on superhydrophobic surfaces
with water/2-propanol solutions of a polyelectrolyte \cite{CR:2006}.

This work was supported  by the National Science Foundation grant
DMR-0808812 and by funds from Syracuse University. We are grateful
for productive discussions with Cristina Marchetti, Pat Mather,
Shiladitya Banerjee and Pine Yang.

{}


\begin{thebibliography}{10}

\bibitem{chem}
F. Brochard,
\newblock {Langmuir } \textbf{5}, 432 (1989).



\bibitem{Q_hot}
D. Qu{\'e}r{\'e} and A. Ajdari,
\newblock {Nat. Mater.} \textbf{5}, 429 (2006).

\bibitem{L_hot}
H. Linke \textit{et al}.,
\newblock {Phys. Rev. Lett.} \textbf{96}, 154502 (2006).


\bibitem{deGennes}
P.G. de Gennes, F. Brochard-Wyart, and D. Qu{\'e}r{\'e},
\newblock {\em Capillarity and Wetting Phenomena: Drops, Bubbles, Pearls,
  Waves}
\newblock (Springer, New York, 2003).


\bibitem{Wenzel}
R.N. Wenzel,
\newblock {Ind. Eng. Chem.} \textbf{28}, 988 (1936).


\bibitem{wrinkles}
D. Vella, M. Adda-Bedia and E. Cerda,
\newblock {Soft Matter} \textbf{6}, 5778 (2010).




\bibitem{cal_1}
The contact area $A_c$ of a droplet of radius $R$ and a sinusoidal
substrate can be calculated in the Cartesian coordinates with the
origin at the center of the droplet. The contact area in the first
quadrant is $A_c/4= \sum_{i=0}^{N=R/\lambda} L_{\lambda}
\sqrt{R^2-x_i^2}$, where $x_i=i \lambda$ $(i=0,1,2.. N)$ are the
positions of the sinusoidal peaks. For $R>> \lambda$,
$\sum_{i=0}^{N=R/\lambda} \triangle i= \int_0^N d i$, where
$\triangle i=1$. $A_c$ can thus be obtained by integration.

\bibitem{Kao}
T. Onda, S. Shibuichi, N. Satoh and K. Tsujii,
\newblock {Langmuir} \textbf{12}, 2125 (1996).


\bibitem{Koch}
H. von Koch,
\newblock {Acta Math.}  \textbf{30}, 145 (1906).



\bibitem{fractal_book}
K. Falconer,
\newblock {\em Fractal Geometry: Mathematical Foundations and Applications} (Wiley, New Jersey, 1990).


\bibitem{NB}
M. Nosonovsky and B. Bhushan,
\newblock {Microsyst. Technol.} \textbf{11}, 535 (2002).

\bibitem{NN}
{\em Nanotribology and Nanomechanics: An Introduction}, edited by
B. Bhushan (Springer, New York, 2005).

\bibitem{Nosonovsky}
M. Nosonovsky,
\newblock {J. Chem. Phys.} \textbf{126}, 224701 (2007).

\bibitem{Landau}
L.D. Landau and E.M. Lifshitz,
\newblock {\em Fluid Mechanics}, 2nd edition
\newblock (Pergamon Press, Oxford, 1987).

\bibitem{PL:2001}
A. R. Parker and C.R. Lawrence,
\newblock{Nature} \textbf{414}, 33 (2001).

\bibitem{CR:2006}
L. Zhai \textit{et al}.,
\newblock{Nano Lett.} \textbf{6}, 1213 (2006).

\end{thebibliography}
\end{document}